\documentclass[10pt]{wlpeerj}

\usepackage[document]{ragged2e}

\usepackage{parskip}
\usepackage{csquotes}
\usepackage{hyperref}
\usepackage[justification=centering]{caption}

\title{Citation method, please? A case study in astrophysics.}

\author{Alice Allen}
\affil{University of Maryland, College Park, MD/Astrophysics Source Code Library}
\corrauthor{Alice Allen}{aallen@ascl.net}

\begin{abstract}
Software citation has accelerated in astrophysics in the past decade, resulting in the field now having multiple trackable ways to cite computational methods. Yet most software authors do not specify how they would like their code to be cited, while others specify a citation method that is not easily tracked (or tracked at all) by most indexers. Two metadata file formats, codemeta.json and CITATION.cff, developed in 2016 and 2017 respectively, are useful for specifying how software should be cited. In 2020, the Astrophysics Source Code Library (ASCL, ascl.net) undertook a year-long effort to generate and send these software metadata files, specific to each computational method, to code authors for editing and inclusion on their code sites. We wanted to answer the question, “Would sending these files to software authors increase adoption of one, the other, or both of these metadata files?” The answer in this case was no. Furthermore, only 41\% of the 135 code sites examined for use of these files had citation information in any form available. The lack of such information creates an obstacle for article authors to provide credit to software creators, thus hindering citation of and recognition for computational contributions to research and the scientists who develop and maintain software.

\end{abstract}

\begin{document}

\flushbottom
\maketitle
\thispagestyle{empty}

\section*{Introduction}
Computational methods underpin the vast majority of today's research \citep{Mushotzky_2011, Goble_2014}, and in the past decade, the landscape of creating, sharing, publishing, and citing scientific research software has changed enormously. Collaborative coding sites such as GitLab,\footnote{\href{https://about.gitlab.com/}{https://about.gitlab.com/}} Bitbucket,\footnote{\href{https://bitbucket.org/product/}{https://bitbucket.org/product/}} and GitHub\footnote{\href{https://github.com/}{https://github.com/}} have made it easier to share the development of software and the source code itself, and archival resources such as Figshare,\footnote{\href{https://figshare.com/}{https://figshare.com/}} Software Heritage,\footnote{\href{https://www.softwareheritage.org/}{https://www.softwareheritage.org/}} and Zenodo\footnote{\href{https://zenodo.org/}{https://zenodo.org/}} offer straightforward paths for preserving research software. Existing journals such as \emph{Science},\footnote{\href{https://www.sciencemag.org/authors/science-journals-editorial-policies\#research-standards}{https://www.sciencemag.org/authors/science-journals-editorial-policies\#research-standards}} \emph{Nature},\footnote{\href{https://www.nature.com/nature-portfolio/editorial-policies/reporting-standards}{https://www.nature.com/nature-portfolio/editorial-policies/reporting-standards}} and the American Astronomical Society Journals\footnote{\href{https://journals.aas.org/news/policy-statement-on-software/}{https://journals.aas.org/news/policy-statement-on-software/}} now request or require software release and citation, and new journals that focus specifically on scientific software have been established \citep{accomazzi_astronomy_2013, beckmann_csbs_2017, smith_joss_2017}. Domain-specific repositories and registries, such as the Astrophysics Source Code Library (ASCL)\footnote{\href{https://ascl.net/}{https://ascl.net/}} and CoMSES Net,\footnote{\href{https://www.comses.net/}{https://www.comses.net/}} make finding this software much easier, and cross-disciplinary efforts, such as FORCE11\footnote{\href{https://www.force11.org/}{https://www.force11.org/}} and the Research Data Alliance FAIR for Research Software (FAIR4RS) Working Group,\footnote{\href{https://www.rd-alliance.org/groups/fair-4-research-software-fair4rs-wg}{https://www.rd-alliance.org/groups/fair-4-research-software-fair4rs-wg}} strive to further improve the ecosystem around research software.

Still, challenges remain. It is often difficult for a researcher to cite code, as many scientific software developers do not state how their software should be cited. Sharing ``how to cite" information on a code's webpage or collaborative coding repository (repo) can mitigate this difficulty. 

Whereas a dozen years ago, there were few ways to cite the use of software \citep{YaleRoundtable_reproducible_2010, howison_herbsleb_2011}, today numerous methods are in use \citep{howison_bullard_2016, collberg_repeatability_2016, soito_citations_2017}, some better than others. Citing an article in which the code has been used or is described (when one could get away with this) was a common way to cite software, and this citation method is still prevalent. This by itself, however, can conflate the purpose of the citation, making it difficult to determine whether a specific citation is for a result in the paper or for use of the software. Furthermore, citing software with only an article does not meet the FORCE11 software citation principles \citep{smith_software_2016}, guidelines that many journals have endorsed and follow. These principles state in part that software should be cited directly, on its own, though the guidelines allow for a describing article to be cited in addition to the software. Footnotes and inline text and links have also appeared in papers as a way to acknowledge software use \citep{howison_bullard_2016}, but these by themselves are not formal citations, do not meet the FORCE11 principles, and should be used only in addition to, not instead of, a formal reference in a paper's bibliography. 

Software developers can help ensure their computational methods are cited by explicitly stating their preferred citation method---what artifact they want used to cite their software---on their code sites. Listing a preferred citation makes clear the expectation that if used, the software should be cited. Proper software citation provides credit to the code author(s), and improves the reproducibility and transparency of research \citep{soito_citations_2017}. As citations are often taken as a proxy for scientific impact \citep{radicchi_universality_2008},  researchers generally prefer that citations for a particular work, result, or software package accrue to one entity rather than spreading the same number of citations over several or many entities, thus specifying what artifact should be used is of value to the software creator. 

Codemeta.json \citep{Jones2017} and CITATION.cff \citep{Druskat2019} are metadata schemas for software. Both of these metadata file types are useful for letting others know how a code should be cited. CITATION.cff files are written in YAML, a ``human friendly data serialization standard for all programming languages.''\footnote{\href{https://yaml.org/}{https://yaml.org/}} These files contain only the information needed for citation. CodeMeta files are rendered in JSON; codemeta.json files can contain information needed to cite the code, and also additional metadata, such as who maintains the software, funding source, and development status, useful to software archives, indexers, and others who want to know more about the software than just how to cite it. Both of these metadata file formats can be read by humans and are machine-actionable. However, the complexity of these files and additional knowledge needed to create them has been a barrier to their adoption. 

To help overcome this barrier, the ASCL provides CITATION.cff and codemeta.json metadata files on demand for its entries; the generated files can then be easily edited by a software author to add metadata the ASCL doesn't provide and placed on the code author's website or repository. The hope is that creating these metadata files for developers can ease and accelerate the adoption of these files. To be clear, ASCL does not generate these metadata files for use on the ASCL; they are provided solely as a convenience for software authors so they do not have to write these metadata files from scratch.

\begin{figure}[ht]
\centering
\frame{\includegraphics[width=0.85\linewidth]{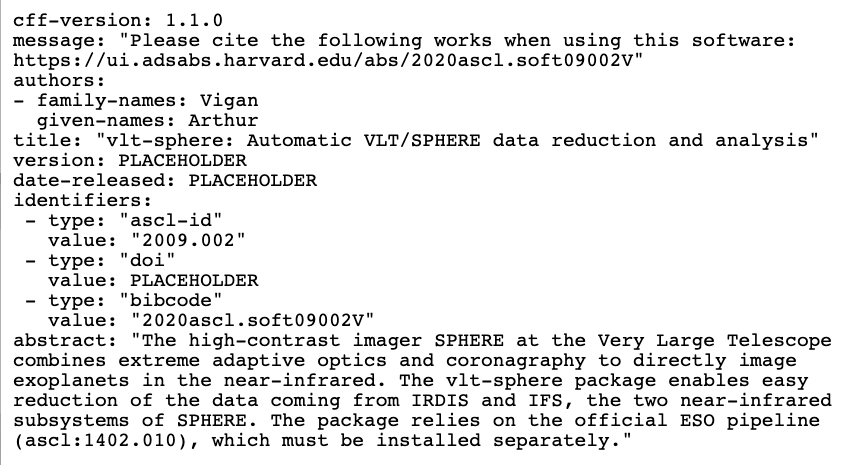}}
\caption{Example of an ASCL-generated CITATION.cff file}
\label{fig:ScreenShot_CITATIONcff_example}
\end{figure}

To create a CITATION.cff file from a ASCL entry (one with an ASCL ID), a user need only add \emph{/CITATION.cff} to the entry's URL, \emph{i.e.}, \url{https://ascl.net/1911.024/CITATION.cff}; the results for this example are shown in Figure~\ref{fig:ScreenShot_CITATIONcff_example}.

\begin{figure}[ht]
\centering
\frame{\includegraphics[width=0.85\linewidth]{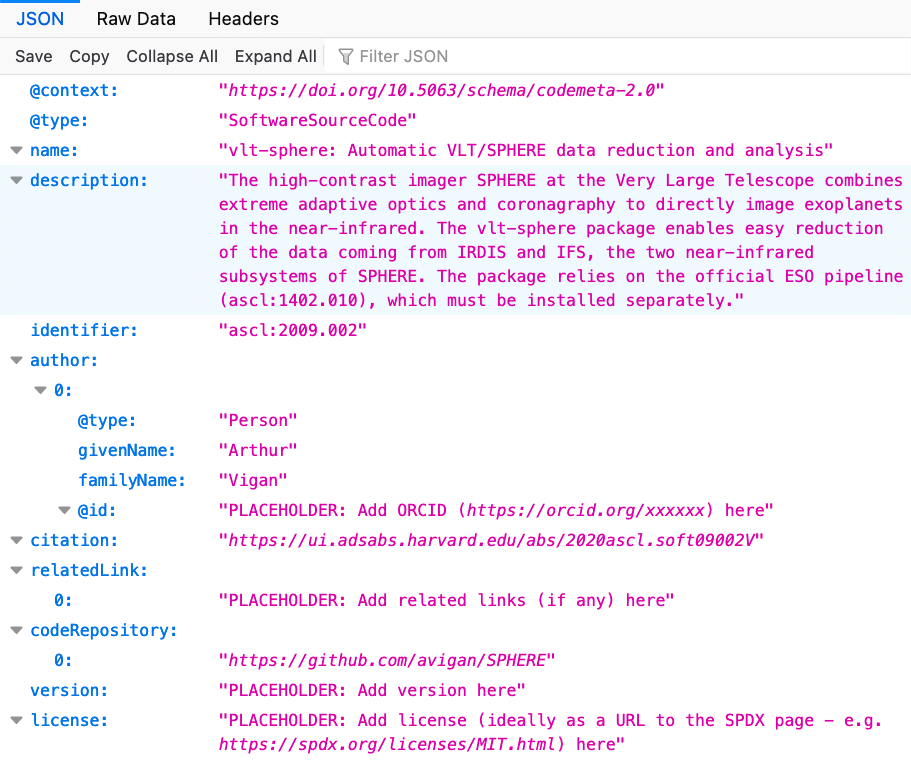}}
\caption{Example of an ASCL-generated codemeta.json file}
\label{fig:ScreenShot_codemetajson_example}
\end{figure}

Similarly, adding \emph{/codemeta.json} to an entry's URL, \emph{i.e.}, \url{https://ascl.net/1911.024/codemeta.json}, will generate a codemeta.json metadata file, as shown in Figure~\ref{fig:ScreenShot_codemetajson_example}.

The ASCL registers scientist-written software that has source code available for immediate download for examination and is used in peer-reviewed astronomy/astrophysics research, in research submitted for peer review, or in research published as an accepted thesis.\footnote{\href{https://ascl.net/home/getwp/1354}{ASCL Editorial policy}} Records for research software are added to the ASCL through submissions by community members, usually but not always authors of the software, and by any of the three ASCL editors, who comb the literature for computational methods. In any given month, new entries are for both old and new software. 

After an entry is created, it is vetted, assigned a unique identifier, the ASCL ID, and put into production. The software package is also downloaded and stored in the ASCL's archive.\footnote{The archive is only for safekeeping; the ASCL does not serve the software it registers to the public unless it has explicit permission from the software author to do so. In general, the ASCL's philosophy is that pointing to the software author's site is preferable than serving the software itself, especially for software under active development and/or frequent use that triggers tweaks and bug fixes.} An ASCL editor then sends a registration notice to each corresponding or lead author of the codes added; each editor corresponds with the authors of the codes she has found. In addition, one editor usually sends the notifications for submitted (not editor-created) entries. This division of labor allows the editors to experiment with the text of the notifications, with one editor varying the standard text, and then, if desired, comparing the results with those of the other editors, whose emails can serve as a control. 

At the end of 2019 and into 2020, one editor inserted text into the registration notices she sent to try to encourage software authors to use the ASCL's codemeta.json and CITATION.cff generator, tweaking the text to try different wordings. A casual look at the results was not encouraging. We then decided to generate and attach either a codemeta.json or CITATION.cff file specific to the code to each email to see whether actually providing the (editable) files would increase use of these files on software sites. It was after this that our real story begins.

\section*{Methods}
Starting in spring 2020 and continuing throughout the year, one editor sent emails for 135 entries to software authors and attached codemeta.json \emph{and} CITATION.cff files specific to the code to each email. These registration notices included information as to the purpose of these files and what to do with them, and were sent for both submitted entries (entries created by a software author) and editor-created entries. The order in which the metadata files were mentioned and attached in the emails was alternated to avoid any issue of implied primacy. Examples of registration notice emails containing these files are in the Appendix.

In January 2021, a list of codes for which these files had been sent was built by searching the ASCL's software archive to see which code directories had both codemeta.json and CITATION.cff files in them, as that indicated which  developers had been sent these files. Of the 135, 133 of the codes had been added to the ASCL in 2020; this represents 52\% of 2020's 257 additions to the ASCL's entries. One of the 135 entries had been added in late December, 2019, and one had been added in 2013.\footnote{For this latter, the metadata files were  generated and sent to the author on March 6, 2020, when an editor emailed the code author to suggest he list a preferred citation method. This code was in the dataset because we'd forgotten we’d done this, and by "we", I mean me.} 

From January 24-26, 2021, the repos/software websites for these 135 codes were examined to see whether a codemeta.json or CITATION.cff file was on the site; if neither of these files was on a code site, we noted whether a preferred citation was designated some other way, or not at all. 

In response to a question as to whether the age of or activity on a code site might be a factor in the uptake of software metadata files, the code sites were examined for last update data (opening tar files when necessary to look for the most recent file date) on February 2, 2021.

On August 19, 2021, the 135 code sites were checked again to see whether there had been a change in the number of codemeta.json or CITATION.cff files on these sites; the sites were not examined for other citation methods on this date.

\section*{Results}
The January 2021 results are shown in Table~\ref{tab:results}.

\begin{table}[ht]
\centering
\begin{tabular}{l|c|r}
Method & Number of code sites & Percentage \\\hline
codemeta.json & 2 & 1.5\% \\
CITATION.cff & 2 & 1.5\% \\
Other & 52 & 38.5\% \\
None & 79 & 58.5\% \\\hline
Total & 135 & 100\% \\
\end{tabular}
\caption{\label{tab:results}Method used to specify preferred citation}
\end{table}

The ``Other" category included citation information in the README file (often supplied in BibTeX or linked to BibTeX), CITATION.md, ACKNOWLEDGEMENT.md, in a PDF on a website (rather than a collaborative coding site), on the software's website, and in a GitHub wiki. Of these other methods, inclusion of citation information in the README file was most prevalent.

There was no change in the number of CITATION.cff or codemeta.json files on the 135 code sites between January and August, 2021.

As mentioned above, we checked the 135 code sites for recent activity. As of February 2, 2021, 70\% of the 135 code sites had been updated after June 30, 2020, 45\% had been updated after October 31, 2020, and 25\% had been updated after December 31, 2020. Seven repos had been updated in February, 2021. 

The data are available in a supplemental .csv file.

\section*{Discussion}
Generating and sending these metadata files did not improve uptake of them. Why? Researchers in general have low familiarity with these metadata file formats; perhaps including the files in an email with other information required too much attention, given that the main purpose of the email was to let authors know their code had been registered or their submission processed. The wording and/or format of the email may be at fault---perhaps using bullet points would have helped---or it could be that these software developers are not familiar with or do not understand the need to and value, for themselves and to other researchers, of specifying a citation method for their code.

With the majority of sites having been updated within the previous twelve months of our examination, we cannot blame the lack of uptake on the age or activity level of the software or repository overall. We had initially intended to compare the results of the emails augmented with metadata files with the code sites/repos of the 48\% of new entry registration notices that did not contain the metadata files (our control group), but the results from sending the metadata files were so low, there was no point in doing so. 

\begin{figure}[ht]
\centering
\frame{\includegraphics[width=0.65\linewidth]{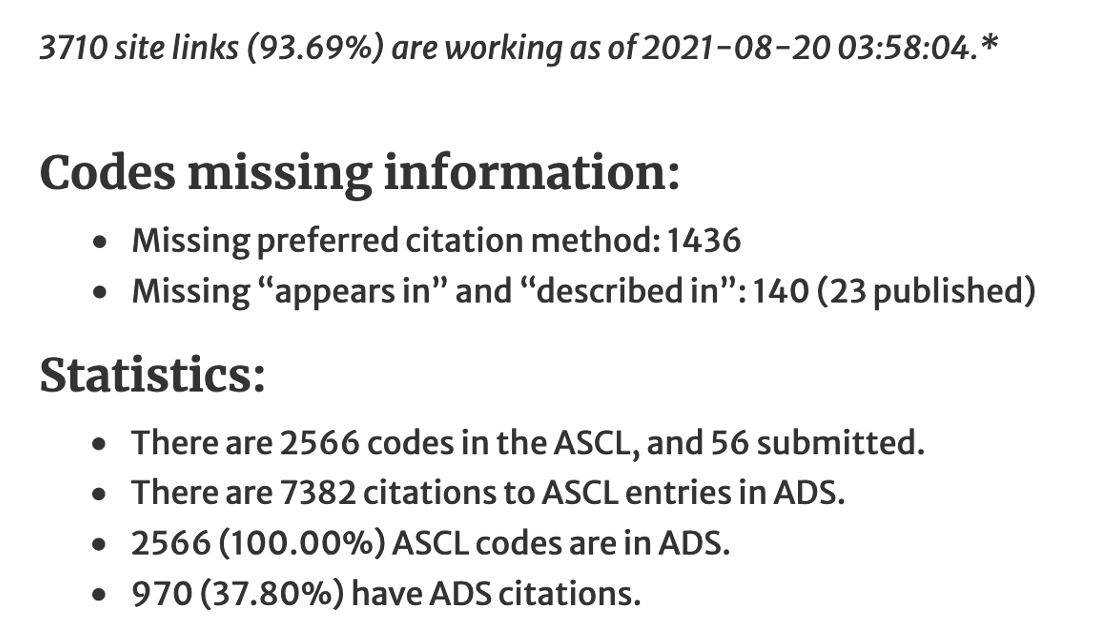}}
\caption{Partial screenshot of ASCL administrator dashboard}
\label{fig:ScreenShot_ASCLDashboard}
\end{figure}

The results do, however, provide us with an estimate as to what percentage of astrophysics code sites do not list the software developer's preferred citation for their work: 58.5\%. As of August 20, 2021, the ASCL’s administrator dashboard, which provides various statistics about the ASCL's entries, shows that preferred citation information is missing for 1,436 entries out of 2,566, which is 56\% of our entries (Figure~\ref{fig:ScreenShot_ASCLDashboard}). The ASCL does not have this information on 56\% of its entries because it is not out there for ASCL editors to find.

Where software metadata should reside, whether on a code site or elsewhere, such as in a registry or repository, is a point of discussion in astrophysics and indeed in the wider scientific community \citep{katz_software_2020, williams_get_2020}. Who might maintain these data has come up in the past as well. In astrophysics, we have the experience of, for example, the now-closed Astrophysics Software and Documentation Service (ASDS) \citep{Hanisch1997}, which expected code authors to maintain the metadata in that resource for their software. This expectation was not met, rendering the service's information increasingly out-of-date and of increasingly less value to the community \citep{allen_looking_2015}. That said, code sites usually already house a good bit of metadata, as at least a partial list of authors for any repo that tracks commits is intrinsically available on the code site, and creating a README file and assigning a license to software are considered best practices \citep{jimenez_four_2017, wilson_good_2017, lee_ten_2018}. 

Recently, GitHub has announced integration of CITATION.cff into its service.\footnote{\href{https://docs.github.com/en/github/creating-cloning-and-archiving-repositories/creating-a-repository-on-github/about-citation-files}{https://docs.github.com/en/github/creating-cloning-and-archiving-repositories/creating-a-repository-on-github/about-citation-files}} This metadata format has also been integrated into Zotero\footnote{\href{https://www.zotero.org/}{https://www.zotero.org/}} and Zenodo \citep{zenodo_2013}; these integrations have resulted in an almost immediate dramatic increase in the number of GitHub repos with CITATION.cff files in their root directories \citep{druskat_2021_integrationslides}. This in turn improves the results that services such as CiteAs \citep{du_citeas_2021} return, making it even easier to find citation information for software.

In addition to the growing integration of CITATION.cff with various services, numerous efforts outside of astronomy also provide metadata files to their software authors using the CodeMeta schema \citep{Alliez_2020, lee_CoMSES_2021}, extract metadata from README.md files and render it in codemeta.json \citep{SOMEF_2019} and further interoperability between other services with this metadata format \citep{habermann_mapping_2019, lamprecht_towards_2020, dicosmo_curated_2020, morrell_2021}. The SciCodes consortium of scientific software registries and repositories\footnote{\href{https://scicodes.net/}{https://scicodes.net/}} is also planning to use CodeMeta to enable interoperability. The ASCL currently offers its public metadata holdings in JSON\footnote{\href{https://ascl.net/code/json}{https://ascl.net/code/json}} and plans to also offer this information in codemeta.json to support interoperability with other services. 

As sending metadata files to authors clearly did not work, and with usage of the CITATION.cff/GitHub integration growing, the ASCL no longer sends these files to authors of the software we register. We do, however, continue to encourage authors to list a preferred citation and have changed our standard registration notice text accordingly. An example of the updated email text is available in the Appendix.

\section*{Conclusions}
The ASCL is always looking for ways to improve its services and processes to support its goals of greater transparency and reproducibility of astrophysics research by making the computational methods used in this research more discoverable. Generating codemeta.json and CITATION.cff files for 135 codes newly registered in the ASCL and emailing these files to the lead developers of these codes fit into our usual workflow fairly easily, and gave us an opportunity to see whether providing these files would help developers over the barrier of having to create a software metadata file from scratch. Though few code authors edited and used the files we sent, the effort wasn't wasted, as it made us look more closely at how much preferred citation information is provided by developers. We discovered that the ASCL has done as good a job as is possible in listing such information in its entries, though still feel frustrated that more than half of our entries do not have preferred citation information. We hope that the new integration of CITATION.cff with GitHub will result in more code authors putting this information on their software sites, thus making it easier for others to give them the credit and acknowledgement they deserve for their computational contributions to research.

\section*{Acknowledgments}
My thanks to Morane Gruenpeter and Neil Chue Hong for interesting questions about our results, P.W. Ryan for creating the metadata files generator, Tracy Huard for formatting suggestions, and Robert Nemiroff, Liz Allen, and an anonymous reviewer, for useful feedback on this article.

\bibliography{citationplease}

\section*{Appendix A: Email examples}

\label{emails}

\begin{figure}
\centering
\frame{\includegraphics[width=0.90\linewidth]{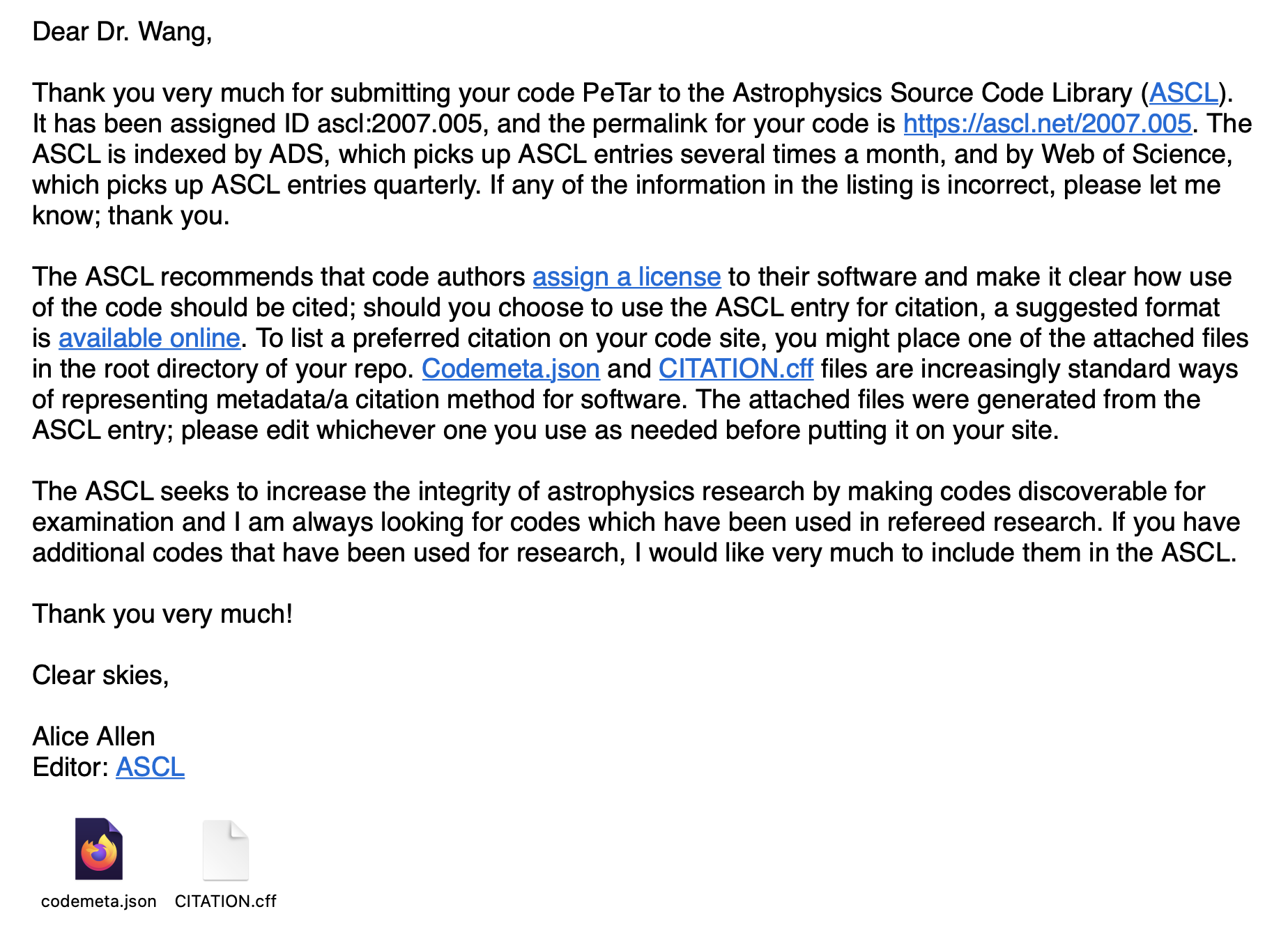}}\\
\raggedright
\raggedright
\caption{Example of a notification email with attached metadata files for a submitted code. The hyperlinks in the second paragraph go (respectively) to the following URLs:\\
\href{https://ascl.net/wordpress/2015/01/05/software-licensing-resources/}{https://ascl.net/wordpress/2015/01/05/software-licensing-resources/}\\
\href{https://ascl.net/home/getwp/351}{https://ascl.net/home/getwp/351}\\
\href{https://codemeta.github.io/index.html}{https://codemeta.github.io/index.html}\\
\href{https://citation-file-format.github.io/}{https://citation-file-format.github.io/}\\}
\label{tab:ScreenShot_submittedemailexample}
\end{figure}

\begin{figure}[!t]
\centering
\frame{\includegraphics[width=0.90\linewidth]{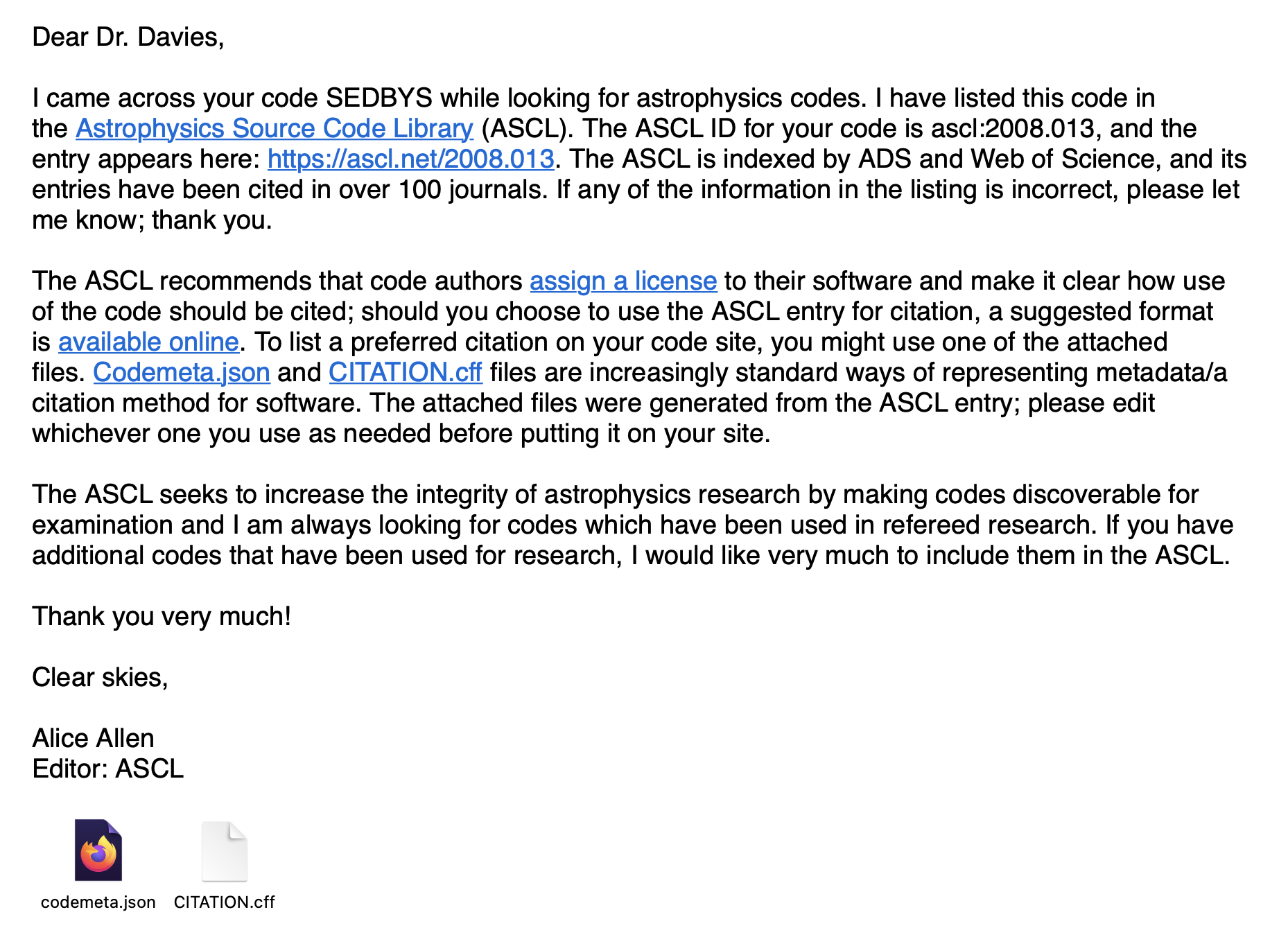}}
\caption{Example of a notification email with attached metadata files for an editor-created ASCL entry. The hyperlinks in the second paragraph go (respectively) to the following URLs:\\
\href{https://ascl.net/wordpress/2015/01/05/software-licensing-resources/}{https://ascl.net/wordpress/2015/01/05/software-licensing-resources/}\\
\href{https://ascl.net/home/getwp/351}{https://ascl.net/home/getwp/351}\\
\href{https://codemeta.github.io/index.html}{https://codemeta.github.io/index.html}\\
\href{https://citation-file-format.github.io/}{https://citation-file-format.github.io/}\\}
\label{tab:ScreenShot_minedemailexample}
\end{figure}

\begin{figure}[ht]
\centering
\frame{\includegraphics[width=0.90\linewidth]{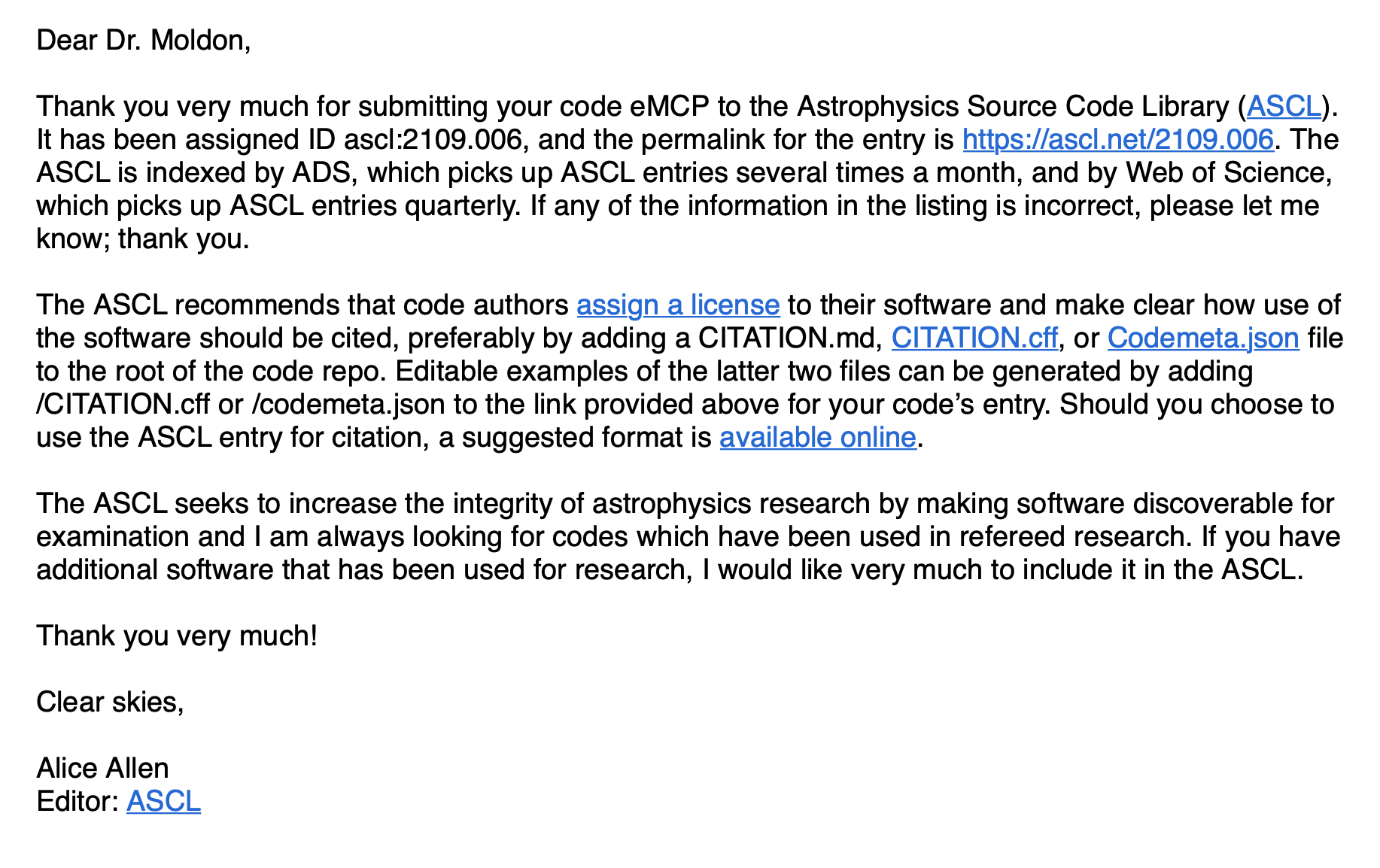}}
\caption{Example of a current notification email for a submitted code; we no longer attach codemeta.json and CITATION.cff files, but do include information about them along with a suggestion to the author to specify his/her preferred citation method. The hyperlinks in the second paragraph go (respectively) to the following URLs:\\
\href{https://ascl.net/wordpress/2015/01/05/software-licensing-resources/}{https://ascl.net/wordpress/2015/01/05/software-licensing-resources/}\\
\href{https://citation-file-format.github.io/}{https://citation-file-format.github.io/}\\
\href{https://codemeta.github.io/index.html}{https://codemeta.github.io/index.html}\\
\href{https://ascl.net/home/getwp/351}{https://ascl.net/home/getwp/351}\\}
\label{tab:ScreenShot_minedemailexample}
\end{figure}

\end{document}